\documentclass[prl,aps,twocolumn]{revtex4}
\usepackage{amsfonts}
\usepackage{amsmath}
\usepackage{amssymb}
\usepackage{bm}
\usepackage{graphicx}
\usepackage{color}

\usepackage{ulem}

\usepackage{epsfig,psfrag,subfigure}%

\begin{document}
\title{Signature of electromagnetic quantum fluctuations  in exciton physics}

\author{Monique Combescot and Fran\c{c}ois Dubin} 
\affiliation{ Institut des Nanosciences de Paris, CNRS and Sorbonne Universit\'e,
75005 Paris, France}

\author{Shiue-Yuan Shiau}
\affiliation{Physics Division, National Center for Theoretical Sciences, Taipei, 10617, Taiwan}

\begin{abstract}
Quantum fluctuations of the electromagnetic field are known to produce the atomic Lamb shift. We here reveal their iconic signature in semiconductor physics,  through  the blue-shift they produce to optically bright excitons, thus lifting the energy of  these excitons above their dark counterparts. The electromagnetic field here acts in its full complexity: in addition to the longitudinal part via interband \textit{virtual  Coulomb} processes, the transverse part---which has been missed up to now---also acts via resonant and nonresonant \textit{virtual photons}. These two parts beautifully combine to produce a bright exciton blue-shift independent of the exciton wave-vector direction. Our work readily leads to a striking prediction: long-lived excitons must have a small bright-dark splitting. Although the analogy between exciton and hydrogen atom could lead us to see the bright exciton shift as a Lamb shift, this is not fully so: the atom shift entirely comes from virtual photons, whereas the Coulomb interaction also contributes to the exciton shift through the so-called ``electron-hole exchange''.
 \end{abstract}
 \date{\today}

\maketitle

Semiconductor excitons \cite{Knox,CHO,Combescot_book} are electron-hole pairs correlated by Coulomb attraction. Due to their internal degrees of freedom, excitons couple differently to light. Since photons have no spin, optically bright excitons are in a spin-singlet state $(S = 0; S_z = 0)$. Importantly, these excitons coexist with optically dark excitons in spin-triplet states $(S = 1; S_z = \pm 1)$ produced by carrier exchange between two bright excitons \cite{Combescot_book,Phys_Rep,Combescot_2015}.

The coexistence of optically bright and dark excitons has numerous consequences on fundamental and applied physics. The seminal experiments on GaAs bulk \cite{Bimberg_1979} and quantum well \cite{Blackwood_94,Amand_97,Lavallard} samples established that dark excitons have a lower energy than bright excitons, despite the fact that electrons and holes are spin-degenerate. As a result, all collective effects that rely on the lowest energy states, like Bose-Einstein condensation \cite{Combescot_2007,Leunberger,Combescot_ROPP,Sean2019}, are inevitably driven by  dark excitons. In the emerging field of Excitonics \cite{Kis_2019}, low energy and poor coupling to photons make dark excitons attractive: in quantum dots, dark states with long lifetime have recently allowed to demonstrate photonic cluster states \cite{Gershoni_2016}. Still, dark states can  play a detrimental role when bright excitons are used for processing optical information in semiconductor devices: the buildup of dark states out of bright states limits the device fidelity and its optical efficiency \cite{Xu_2018,Robert_2020}.

So far, the bright-dark exciton splitting has been ascribed to the so-called ``electron-hole exchange'' through ``short-range'' and ``long-range'' Coulomb processes \cite{Pikus_1971,Sham_93,Fu_99,Luo2009}. However, this understanding based on Coulomb interaction only, surely misses part of the physics because the splitting it produces cancels for some directions of the exciton wave vector (see Eq.~(\ref{7})), a physically odd result that has never been questioned. The missed physics comes from the virtual photons that bright excitons can emit and reabsorb, as in the atomic Lamb shift \cite{Lamb_47, Bethe_47,Cohen_Atom_Photon}, thereby inducing a similar blue shift. This urges the physics behind the bright-dark exciton splitting to be reconsidered from scratch, as a first step toward controlling and manipulating these two types of excitons in novel experiments and semiconductor devices.

\textbf{In this Letter, we establish} the microscopic origin of the splitting between dark and bright excitons and relate this splitting to the quantum fluctuations of the longitudinal and transverse electromagnetic fields.

 We show that the bright exciton blue-shift  not only comes from  Coulomb interaction through virtual interband processes, commonly called ``electron-hole exchange'', but also from processes involving virtual photons. Since dark excitons suffer neither of these processes, their energy stays unchanged; so, the bright-dark  exciton splitting is equal to the bright exciton energy shift. As Coulomb interaction follows from the longitudinal part of the electromagnetic field, while photons are associated with the transverse part, the bright-dark exciton splitting constitutes an iconic signature, in semiconductor physics, of quantum fluctuations in the theory of radiation, as, \textit{mutatis mutandis}, does the Lamb shift in atomic physics.

What sets excitons apart from hydrogen atoms is that the electromagnetic field  acts in two totally different ways: the longitudinal field acts through \textit{inter}band Coulomb processes, while the transverse field acts through ``resonant'' and ``nonresonant'' \textit{inter}band processes, that is, photon absorption along with exciton creation and recombination. The nonresonant coupling, commonly dropped for physics driven by real photons, must be kept for processes involving virtual photons: as a mathematical proof, the longitudinal and transverse electromagnetic fields produce contributions that respectively depend on the longitudinal and transverse components of the exciton wave vector $\textbf{K}$, which is physically reasonable. Their contributions nicely combine to produce an energy shift free from these components, hence supporting that bright excitons do feel electromagnetic field fluctuations in their full complexity.  

\begin{figure}
\includegraphics[width=.4\textwidth]{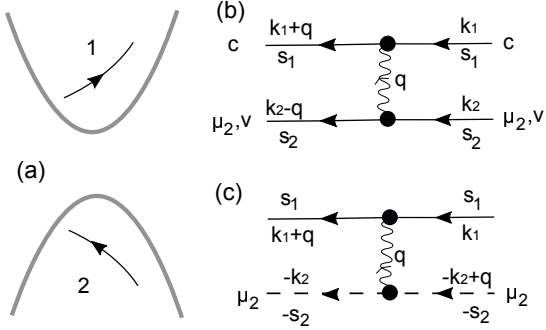}
\caption{(a) \textit{Intra}band Coulomb interaction. Diagrams in terms of (b) conduction and valence electrons and (c) electrons and holes. Electrons are represented by solid lines and holes by dashed lines. The spin $s$ and valence spatial index $\mu=(x,y,z)$ of semiconductor Bloch states are conserved in intraband Coulomb processes.}\label{fig1}
\end{figure}

\textbf{We moreover predict} that the energy splitting $\Delta_{BD}$  between bright and dark excitons must scale as 
\begin{equation}
\label{1}
\Delta_{BD}\propto \frac{\,\,|\Omega_{ph-X}|^2}{E_{gap}}
\end{equation}
 where the energy-like parameter $\Omega_{ph-X}$ is the Rabi coupling between exciton and photon, and $E_{gap}$ is the band gap. As a direct consequence, the bright-dark splitting  varies as the inverse of the exciton radiative lifetime. Although here derived for cubic crystals, this relation stays valid whatever the material structure and shape \cite{note1}, as supported by a strong dimensional argument:  the physics of the shift  involves two exciton-photon couplings; to end with an energy-like shift, they must be divided by an energy, which can only be the energy that creates the exciton, that is, the band gap.
 This amazingly simple result will provide a useful tool for characterizing optically inactive excitons in new materials.

\noindent \textbf{(1)}\textbf{\textit{ Couplings to the electromagnetic field}}

$\bullet$ The Coulomb interaction comes from the longitudinal part of the electromagnetic field. There are two types of Coulomb processes. 

\noindent \textit{(i)} \textit{Intra}band processes (Fig.~\ref{fig1}) correlate electron-hole pair into exciton. Since Coulomb interaction does not act on spin, each carrier keeps its spin. Moreover, the hole keeps its spatial index $\mu$ due to spatial symmetry \cite{Combescot_book}.  

\begin{figure}
\includegraphics[width=.45\textwidth]{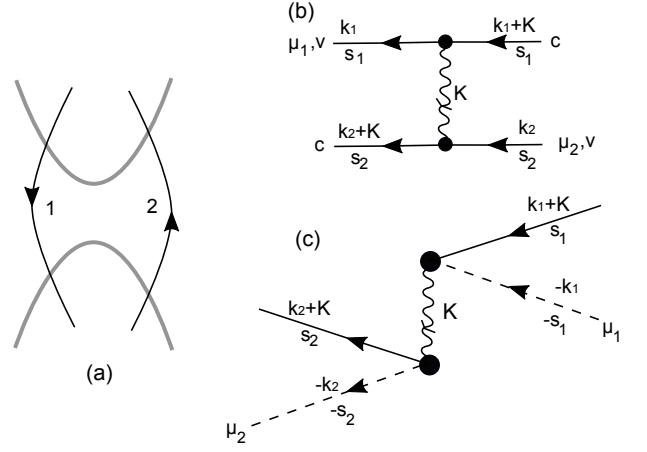}
\caption{(a) \textit{Inter}band Coulomb interaction. Diagrams in terms of (b) conduction and valence electrons and (c) electrons and holes. The associated vertex is given by Eq.~(\ref{2}). The wave-vector transfer \textbf{K} is equal to the pair center-of-mass wave vector.}
\label{fig2}
\end{figure}

\noindent  \textit{(ii)}  \textit{Inter}band  processes (Fig.~\ref{fig2})  correspond to the recombination of an electron-hole pair and the excitation of another pair. The fact that Coulomb interaction conserves the spin imposes the electron-hole pairs that experience interband processes to be in a spin-singlet state \cite{Combescot_book}. The interband Coulomb vertex 
 depends on the center-of-mass wave vector \textbf{K} of the pairs as \cite{SM} 
 \begin{equation}\label{2}
 \frac{2|\Lambda|^2}{E_{gap}^2} \Big( \frac{\textbf{K}}{K}\cdot\textbf{e}_{\mu_1} \Big) \Big( \frac{\textbf{K}}{K}\cdot\textbf{e}_{\mu_2} \Big)
 \end{equation}
 for pairs made of hole in a threefold spatial state $\mu$. The unit vector $\textbf{e}_\mu$ with $\mu=(x,y,z)$ is along a cubic crystal axis, while the $\Lambda$ parameter is defined in Eq.~(\ref{4}). In contrast to \textit{intra}band scatterings, the hole indices  $(\mu_1,\mu_2)$ of the ``in'' and ``out'' pairs are not related (Fig.~\ref{fig2}(c)).

\begin{figure}
\includegraphics[width=.48\textwidth]{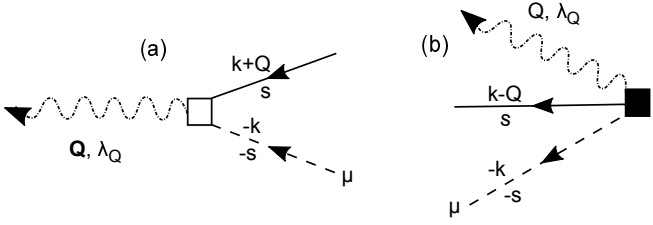}
\caption {\textit{Inter}band electron-photon interaction. (a) Resonant process: the photon $\textbf{Q}$ creation goes along with the recombination of an electron-hole pair with center-of-mass wave vector $\textbf{Q}$; (b) nonresonant process: the photon $\textbf{Q}$ creation  goes along with the creation of an electron-hole pair with center-of-mass wave vector $-\textbf{Q}$. Photons are represented by dotted wavy  lines. The associated vertices are given in Eq.~(\ref{3}).}\label{fig3}
\end{figure}

$\bullet$ Photons associated with the transverse field also lead to \textit{inter}band processes. The emission of a photon with wave vector \textbf{Q} goes along with the recombination  of an electron-hole pair with wave vector \textbf{Q}  (resonant process, Fig.~\ref{fig3}(a)), or the creation of a pair with wave vector $-\textbf{Q}$  (nonresonant process, Fig.~\ref{fig3}(b)). Since photons have no spin, the involved pair must be spin-singlet, the photon polarization vector 
 $\textbf{e}_{\lambda,\textbf{Q}}$ with $\lambda=(X,Y)$, orthogonal to  $\textbf{e}_{Z,\textbf{Q}}=\textbf{Q}/Q$, acting on the hole wave function. The vertex for photon with energy $\hbar\omega_\textbf{Q}$ and spin-singlet pair with a $\mu$ hole, reads \cite{SM}
 \begin{equation}
 \label{3}
 \frac{\Lambda}{\sqrt{\hbar\, \omega_\textbf{Q}}}              \Big   (\textbf{e}_{\lambda,\textbf{Q}}\cdot\textbf{e}_{\mu}   \Big )
 \end{equation}
 for resonant processes, $\Lambda$ being replaced by $\Lambda^*$ for non-resonant processes.

$\bullet$ As salient features of these couplings: 

\noindent\textit{(i)} the interband Coulomb interaction and the electron-photon interaction both act on spin-singlet pairs, which are bright; 

\noindent\textit{(ii)} although Coulomb interaction and  electron-photon interaction follow \cite{Combescot_book} from different physics, ($e^2/r$) and $(-q_e\textbf{A}\cdot \hat{\textbf{p}}/ m_0)$, their interband vertices read in terms of the same parameter $\Lambda$ that appears in CGS and MKS units as
  \begin{equation}
   \label{4}
  \Lambda=\frac{\hbar P_{c,v}}{m_0}\sqrt{\frac{4\pi e^2}{L^3}}=\frac{q_eP_{c,v}}{m_0}\sqrt{\frac{\hbar^2}{\epsilon_0L^3}}
   \end{equation}
for free electron mass $m_0$ and sample volume $L^3$.
The momentum operator $\hat{\textbf{p}}$ between conduction and valence Bloch states taken at the $(\textbf{k}=\textbf{0})$ band extremum, $P_{c,v}= \langle u_{c;\textbf{0}} |  \textbf{e}_\mu\cdot \hat{\textbf{p} }| u_{\mu,v;\textbf{0}}\rangle$, is  $\mu$ independent for cubic crystals.

 
   
   
   
\noindent \textbf{(2)}\textbf{\textit{ Bright exciton shift}}

$\bullet$ The repetition of \textit{intra}band Coulomb processes transforms a spin-singlet free pair made of a $\mu$ hole and having a center-of-mass wave vector \textbf{K} into a bright exciton, $|X_{\textbf{K}, \nu;\mu,S=0}\rangle$, its energy $E_{\textbf{K}, \nu}$ being $\mu$  independent; the exciton relative-motion index $\nu$ will be denoted as $\nu_0$ for ground state.

\begin{figure}
\includegraphics[width=.48\textwidth]{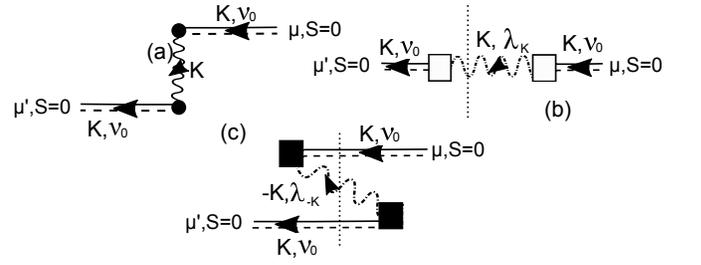}
\caption{(a) Interband Coulomb interaction producing the energy shift Eq.~(\ref{7}) for exciton with wave vector \textbf{K}. (b,c) Interband electron-photon interaction: resonant processes (b) lead to the energy shift Eq.~(\ref{9}), while nonresonant processes (c) lead to the shift Eq.~(\ref{10}). The intermediate states (indicated by vertical dotted lines) correspond to a virtual photon \textbf{K} in (b) and a virtual photon $-$\textbf{K} plus two excitons in (c).
}\label{fig4}
\end{figure}

$\bullet$ The exciton energy change induced by \textit{inter}band Coulomb interaction follows from Eq.~(\ref{2}). The matrix elements of this interaction in the $\mu$ degenerate ground-exciton subspace read \cite{SM} 
     \begin{eqnarray}
      \label{7}
 \Delta_{Coul}^{( \mu',\mu)}&=&  \langle X_{\textbf{K}, \nu_0;\mu',S=0} |\hat{V}_{Coul}^{(inter)}|  X_{\textbf{K}, \nu_0;\mu,S=0}\rangle
\nonumber
 \\
 &=&    \frac{2|\Lambda_{\nu_0}|^2}{E_{gap}^2}  \Big( \frac{\textbf{K}}{K}\cdot\textbf{e}_{\mu'} \Big)
 \Big( \frac{\textbf{K}}{K}\cdot\textbf{e}_{\mu} \Big)
   \end{eqnarray}
   with $\Lambda_{\nu_0}=\Lambda  \,L^{3/2} \,  \langle \textbf{r}=\textbf{0} |\nu_0\rangle$: 
  a $\mu$ exciton recombines while a $\mu'$ exciton is created (Fig.~\ref{fig4}(a)). As $ \Delta_{Coul}^{( \mu',\mu)}$ cancels for \textbf{K} orthogonal to a crystal axis, an additional process is required to have a nonzero shift whatever \textbf{K}.
   

    
 $\bullet$ A similar energy change follows from the interband electron-photon interaction $\hat{W}_{ph}^{(inter)}$ through the emission and reabsorption of virtual photons. The resulting shift, quadratic in $\hat{W}_{ph}^{(inter)}$, appears as
 \begin{eqnarray}
\Delta_{ph}^{( \mu',\mu)}=
\,\,\,\, \,\,\,\,\,\,\,\,\,\,\,\,\,\,\,\,\,\,\,\,\,\,\,\,\,\,\,\,\,\,\,\,\,\,\,\,\,\,\,\,\,\,\,\,\,\,\,\,\,\,\,\,\,\,\,\,\,\,\,\,\,\,\,\,\,\,\,\,\,\,\,\,\,\,\,\,\,\,\,\,\,\,\,\,\,\,\,\,\,\,\,\,\,\,\,\,\,\,\,\,\,\,\,\,\,\,\,\,\,\,\,\,\,\,\,\,
 \\
 \sum_f\frac{\langle X_{\textbf{K}, \nu_0;\mu',S=0}| \hat{W}_{ph}^{(inter)} |f \rangle \langle  f| \hat{W}_{ph}^{(inter)} | X_{\textbf{K}, \nu_0;\mu,S=0} \rangle}{E_{\textbf{K}, \nu_0} - E_f}
 \nonumber
  \end{eqnarray} 

\noindent \textit{(i)} For resonant processes, photon emission goes along with exciton recombination; so, the intermediate state $|f\rangle$ is just the virtual photon \textbf{K} (Fig.~\ref{fig4}(b)) with energy $E_f=\hbar\, \omega_\textbf{K}$.  Equation (\ref{3}) then gives 
   \begin{eqnarray}
     \label{9}
   \Delta^{( \mu',\mu)}_{res}=
   \Big | \frac{\Lambda_{\nu_0}}
  {\sqrt{\hbar\, \omega_\textbf{K}}} 
  \Big |^2  
   \frac   {1}  {E_{\textbf{K},\nu_0}-\hbar\omega_\textbf{K}}
   \,\,\,\,   \,\,\,\,  \,\,\,\,  \,\,\,\,  \,\,\,\,  \,\,\,\,  \,\,\,\, 
 \\
  \nonumber   
      \sum_{\lambda=(X,Y)}
       \Big(\textbf{e}_{\lambda,\textbf{K}}  \cdot\textbf{e}_{\mu'} \Big)
        \Big(\textbf{e}_{\lambda,\textbf{K}}  \cdot\textbf{e}_{\mu} \Big) 
   \end{eqnarray}
  
\noindent \textit{(ii)} For nonresonant processes, photon emission goes along with exciton creation; the
intermediate state $|f\rangle$ is made of a virtual photon $-\textbf{K}$ plus two excitons (Fig.~\ref{fig4}(c)); so, $E_f =\hbar\, \omega_{-\textbf{K}}+2E_{\textbf{K},\nu_0}$ and Eq.~(\ref{3}) give
     \begin{eqnarray}
       \label{10}
   \Delta^{( \mu',\mu)}_{nres}&=&  \Big | \frac{\Lambda_{\nu_0}}
  {\sqrt{\hbar\, \omega_\textbf{K}}} 
  \Big |^2           
       \frac   {1}  {-E_{\textbf{K},\nu_0}-\hbar \omega_{-\textbf{K}}} 
       \\
  \nonumber     
         && \sum_{\lambda=(X,Y)}               \Big(\textbf{e}_{\lambda,\textbf{K}}  \cdot\textbf{e}_{\mu'} \Big)
                              \Big(\textbf{e}_{\lambda,\textbf{K}}  \cdot\textbf{e}_{\mu} \Big)
      \end{eqnarray}

 Combining the resonant and nonresonant contributions yields, since $\omega_\textbf{K}=\omega_{-\textbf{K}}$,
        \begin{equation}\label{11_1}
   \Delta_{ph}^{( \mu',\mu)}=   \frac{2|\Lambda_{\nu_0}|^2}{E_{\textbf{K},\nu_0}^2 - \hbar^2\omega_\textbf{K}^2} 
  \sum_{\lambda=(X,Y)} \Big(\textbf{e}_{\lambda,\textbf{K}}  \cdot\textbf{e}_{\mu'} \Big)
    \Big(\textbf{e}_{\lambda,\textbf{K}}  \cdot\textbf{e}_{\mu} \Big)
      \end{equation}
with $\hbar\omega_\textbf{K} \ll E_{\textbf{K},\nu_0}\simeq E_{gap}$, as $|\textbf{K}|\simeq 0$ for photon.


$\bullet$ From Eqs.~(\ref{7},\ref{11_1}), we see that for $\textbf{K}\parallel\textbf{e}_{\mu}$, the contribution from virtual photons cancels while the one from interband Coulomb interaction is maximum; this is the opposite for $\textbf{K} \bot \textbf{e}_{\mu}$, as reasonable because the Coulomb interaction comes from the longitudinal field, while photons are associated with the transverse field.
 Actually, these contributions nicely combine: as $(\textbf{e}_{X,\textbf{K}},\textbf{e}_{Y,\textbf{K}})$ and $\textbf{e}_{Z,\textbf{K}}=\textbf{K}/K$ form an orthonormal set,
 \begin{eqnarray}
\delta_{\mu',\mu}&=&\textbf{e}_{\mu'} \cdot \textbf{e}_{\mu}\nonumber
\\
&=&\!\!\sum_{\lambda'=(X,Y,Z)}\!\!\!  ( \textbf{e}_{\mu'} \cdot \textbf{e}_{\lambda',\textbf{K}})  \textbf{e}_{\lambda',\textbf{K}}
\cdot
\!\!\! \sum_{\lambda=(X,Y,Z)} \!\!\!  ( \textbf{e}_{\mu} \cdot \textbf{e}_{\lambda,\textbf{K}})  \textbf{e}_{\lambda,\textbf{K}}
\nonumber\\
&=&\!\!\! \sum_{\lambda=(X,Y,Z)}\!\!\!  \left( \textbf{e}_  {\mu'} \cdot \textbf{e}_{\lambda,\textbf{K}} \right)
\left(\textbf{e}_\mu  \cdot\textbf{e}_{\lambda,\textbf{K}} \right)
\end{eqnarray}
 which proves that the sum of the longitudinal and transverse shifts does not depend on the $\textbf{K}$ direction 
         \begin{equation}
    \label{11}     
 \Delta^{(\mu',\mu)}= \Delta_{Coul}^{(\mu',\mu)}+  \Delta_{ph}^{(\mu',\mu)}\simeq \delta_{\mu',\mu} \frac{2|\Lambda_{\nu_0}|^2}{E_{gap}^2 }   
      \end{equation}
      
 This bright exciton shift corresponds to the splitting between bright and dark excitons given in Eq.~(\ref{1}):   
 
 \noindent\textit{(i)} dark excitons are not coupled to the electromagnetic field; so, their energy does not change; 
 
 \noindent\textit{(ii)} according to Eq.~(\ref{3}), the Rabi coupling for photon that creates a ground exciton, $\hbar\omega_\textbf{K}\simeq E_{gap}$, scales as 
 \begin{equation}
\Omega_{ph-X}\propto  \frac{\Lambda_{\nu_0} }{\sqrt{E_{gap}}}
 \end{equation}
. 
         
\noindent\textit{(iii)} as the shift does not depend on the hole index $\mu$, the spin-orbit interaction, which enforces specific linear combinations of hole states, can only change the numerical prefactor of the equation (\ref{11}),  but not its form \cite{SOcase}.

\noindent \textbf{(3) \textit{Discussion}}

$\bullet$ Through Eq.~(\ref{1}), the dark-bright exciton splitting is related to two fundamental quantities: the band gap $E_{gap}$ and the  exciton-photon Rabi coupling that depends on the bulk exciton Bohr radius $a_{_{X}}$ and the momentum operator between valence and conduction states, $P_{c,v}$. Using Eq.~(\ref{4}) and the ground-exciton wave function $e^{-r/a_{_{X}}} / \sqrt{\pi a_{_{X}}^3}$, we get \cite{dielectric}
  \begin{equation}
  \Lambda_{\nu_0}=\Lambda  \,L^{3/2} \,  \langle \textbf{r}=\textbf{0} |\nu_0\rangle=\Lambda\sqrt{ \frac{    L^3   }{ \pi a_{_{X}}^3      } }=2\frac{ e  \hbar  P_{c,v} } {m_0 a_{_{X}}^ {3/2}\sqrt{\epsilon_{sc}}}\label{13}
  \end{equation}
  Equation (\ref{11}) then gives the bright-dark splitting as
    \begin{equation} 
    \label{14}            
    \Delta_{BD}=2\frac {|\Lambda_{\nu_0}|^2} {E_{gap}^2}  = 8\, \frac {E_{Kane}} {E_{gap}} \,  \frac {e^2/\epsilon_{sc}a_{_{X}}} {E_{gap}}   \,   \frac {  \hbar ^2} {2m_0 a_{_{X}}^2}
     \end{equation}
 with the ``Kane energy'' $E_{Kane}=2|P_{c,v}|^2/m_0$.
 
 For bulk GaAs ($E_{gap}\simeq1.5$ eV, $E_{Kane}\simeq23$ eV, $\epsilon_{sc}\simeq 13$ and $a_{_{X}}\simeq15$ nm \cite{Bastard}), the above equation gives a $\Delta_{BD}\sim100$ $\mu$eV splitting which is somewhat larger than the reported value 20 $\mu$eV\cite{Bimberg_1979}. As possible reasons: \textit{(i)} the experimental value has not been measured directly, but deduced from comparing computed and measured magneto-luminescence spectra; \textit{(ii)} $E_{Kane}$ and $a_{_{X}}^3$ are not known with high precision; \textit{(iii)} constructing the exciton on spin-orbit eigenstates will change the numerical prefactor in Eq.~(\ref{13}). We hope that the present Letter will stimulate more experiments on bright-dark exciton splittings.

$\bullet$ Although  derived  for bulk crystals, the transparent physics expressed in Eq.~(\ref{1}) leads us to embrace its validity for reduced dimensions. We first note that $a_{_{X}}^3$ in Eq.~(\ref{14}) comes from the exciton volume, $4 \pi a_{_{X}}^3 /3$, over which the Coulomb or photon-exciton interaction occurs. In quantum well, this volume is $w\,\pi a_{_{X}}^2(w)$, with the well width $w$ by construction smaller than $a_{_{X}}$. The in-plane Bohr radius $a_{_{X}}(w)$ is also smaller as it decreases from $a_{_{X}}$ to $a_{_{X}}/2$ when $w$ decreases from infinity to zero. The exciton volume thus is smaller for quantum well than for bulk, therefore making the bright-dark exciton splitting larger, as seen from their ratio for bulk and quantum well, estimated through
 \begin{eqnarray} 
    \label{15}            
   \left(\frac {4\pi} {3} a_{_{X}}^3\right) \,\,  \Delta_{BD}\sim \Big(w \, \pi a_{_{X}}^2(w)\Big) \,\,  \Delta_{BD}^{\rm{(QW)}} 
     \end{eqnarray}
  For $w=5$ nm and $a_{_{X}}(w)\simeq 0.7~a_{_{X}}$  \cite{Bastard}, this gives $\Delta_{BD}^{\rm{(QW)}}/ \Delta_{BD}\sim7$, in agreement with experimental data for GaAs bulk \cite{Bimberg_1979} and quantum well \cite{Blackwood_94,Amand_97} samples.

$\bullet$ Another bright-dark splitting of interest is for interlayer excitons in coupled GaAs quantum well, which have been used to explore Bose-Einstein condensation and related superfluidity \cite{Combescot_ROPP,Beian_2017,Rapaport_trap,Alloing_2014,Anankine_2017,Dang_2019,High_2012}. We can estimate the bright-dark splitting by noting that it varies as the inverse of the exciton radiative lifetime. So, starting from the splitting measured for intralayer excitons in a single GaAs quantum well \cite{Blackwood_94,Amand_97}, we can derive the interlayer splitting from the change in radiative lifetime \cite{MFF}. Experiments have shown that the radiative lifetime for \textit{intra}layer excitons is  a few tens of picoseconds \cite{Deveaud_91}, while it is three orders of magnitude larger \cite{Sivalertpron_2012,Beian_2017} for interlayer excitons. This gives the bright-dark splitting for coupled GaAs quantum well in the range of 200 neV. Interestingly, this value is comparable to the energy difference between excitons having the two lowest center-of-mass wave vectors $0$ and $(2\pi/L)$, for in-plane size $L$ of a few $\mu$m. The $\Delta_{BD}$ amplitude thus confirms the dominant role of optically dark states for exciton Bose-Einstein condensation reported in coupled GaAs quantum wells \cite{Alloing_2014,Anankine_2017,Dang_2019}.

$\bullet$ The physical understanding we provide for the bright exciton blue-shift has some connection with the Lamb shift \cite{Lamb_47} between the 2S$_{1/2}$ and 2P$_{1/2}$ levels of a hydrogen atom, which results from the emission and reabsorption of virtual photons \cite{Bethe_47,Cohen_Atom_Photon}.  Yet, although semiconductor excitons are commonly seen as the solid-state analog of hydrogen atoms, their couplings to the electromagnetic field are qualitatively different due to wave-vector conservation, since holes are much lighter than proton. Indeed, when a photon is emitted by the recombination of an exciton in a direct-gap semiconductor, the photon is given not only the energy but also the center-of-mass wave vector of the exciton \cite{Combescot_book}.  By contrast, the photon emitted by an atom results from the electron decay between allowed states, which determines the emitted photon energy but not its wave vector due to the heavy-atom recoil \cite{Cohen_Atom_Photon}.

$\bullet$ Finally, the physics we here present heavily relies on the difference between exciton and polariton. 

\noindent---In the polariton regime, the exciton-photon coupling is ``strong'' and must be included exactly; strong actually means large compared with the inverse carrier coherence time. For short coherence time, the exciton changes its wave vector faster than photon reemission; the original photon is lost, and the polariton mode can not develop.

\noindent---In the exciton regime, the exciton-photon coupling is ``weak'' and  can be treated at lowest order, as here done. It produces the long-missed part of the bright exciton shift, which is necessary to produce a bright-dark splitting that does not cancel for some exciton wave-vector directions, as physically reasonable.

  \noindent\textbf{\textit{Conclusion}}
 
 We have analytically derived the energy splitting between optically bright and dark excitons in cubic crystals, using the second quantization formalism for field and matter, as required when dealing with quantum fluctuations. We trace this splitting back to the Coulomb interaction through interband virtual processes, joined by the electron-photon interaction through interband resonant and nonresonant virtual processes, these two interactions being related to the longitudinal and transverse parts of the electromagnetic field. Their contributions increase the bright exciton energy, but do not affect dark excitons. As physically reasonable, the bright-dark splitting  does not depend on the direction of the exciton wave vector, a point that would have been missed by considering only Coulomb interaction through ``electron-hole exchange'', as previously done \cite{Pikus_1971,Sham_93,Fu_99,Luo2009}, or only virtual photon emission and absorption, as for the atomic Lamb shift \cite{Lamb_47, Bethe_47,Cohen_Atom_Photon}. Our key result, Eq.~(\ref{1}), leads us to predict that long-lived excitons, weakly coupled to photons, must have a small bright-dark energy splitting, a valuable result in view of the difficulty in measuring quantities related to dark states.
 
 \textbf{Acknowledgments:} We thank T. Amand and X. Marie  for critical inputs regarding the numerical evaluation of the bright-dark splitting in bulk GaAs. Our work has been financially supported by the Labex Matisse and by OBELIX from the french Agency for Research (ANR-15-CE30-0020).

\newpage

\setcounter{figure}{0}
\renewcommand{\figurename}{Figure S\!}

\renewcommand{\theequation}{\mbox{S.\arabic{equation}}} 
\setcounter{equation}{0}

\textbf{Supplementary informations}


\section{Coulomb interaction}

$\bullet$  The Coulomb interaction 
\begin{equation}
 \frac{1}{2}\sum_j\sum_{j'\not=j'} \frac{e^2}{|\textbf{r}_j-\textbf{r}_{j'}|}
 \end{equation}
  written in second quantization on the Bloch-state basis with wave function
\begin{equation}\label{app:1}
\langle \textbf{r}|n,\textbf{k} \rangle=\frac{e^{i\textbf{k}\cdot \textbf{r}}}{L^{3/2}} u_{n,\textbf{k}}(\textbf{r})
\equiv e^{i\textbf{k}\cdot \textbf{r}} \langle \textbf{r}|u_{n,\textbf{k}} \rangle
 \end{equation}
  reduces for small momentum transfer to \cite{appCombescot_book} 
\begin{align}
\hat{V}_{Coul}\simeq \frac{1}{2}
{\sum_{\textbf{q},\{n,\textbf{k},s\}}} 
\mathcal{V}(n'_1,\textbf{k}_1{+}\textbf{q};n_1,\textbf{k}_1)\mathcal{V}(n'_2,\textbf{k}_2;n_2,\textbf{k}_2{+}\textbf{q})  
\nonumber\\
   \hat{a}^\dag_{n'_1,\textbf{k}_1{+}\textbf{q},s_1}\hat{a}^\dag_{n'_2,\textbf{k}_2,s_2}
   \hat{a}_{n_2,\textbf{k}_2{+}\textbf{q},s_2}\hat{a}_{n_1,\textbf{k}_1,s_1}\label{app:2}
\end{align}
with $\mathcal{V}(n',\textbf{k}';n,\textbf{k})=\langle u_{n',\textbf{k}'}|u_{n,\textbf{k}} \rangle\sqrt{4\pi e^2/L^3|\textbf{k}'-\textbf{k}|^2}$. The physically relevant band indices $n$ are $c$ for conduction state and $(v,\mu )$ for threefold valence states labeled as $\mu=(x,y,z)$ with $\textbf{e}_\mu$ along the crystal axes. 

$\bullet$ The scalar product of  Bloch states, $\langle u_{n',\textbf{k}'}|u_{n,\textbf{k}} \rangle$, is obtained in the small $\textbf{q}$ limit, from
\begin{equation}\label{app:3}
|u_{n',\textbf{k}{+}\textbf{q}}\rangle {\simeq} |u_{n',\textbf{k}}\rangle{+}\frac{\hbar \textbf{q}}{m_0}\cdot\sum_{n\neq n'}|u_{n,\textbf{k}}\rangle\frac{\langle u_{n,\textbf{k}}|\hat{\textbf{p}}| u_{n',\textbf{k}}\rangle }{\varepsilon_{n',\textbf{k}}-\varepsilon_{n,\textbf{k}}}{+}\mathcal{O}(\textbf{q}^2)
\end{equation}
where $m_0$ is the free electron mass, the scalar product reducing to zero for $(n,n')=(v,\mu)$ due to symmetry.

\subsection{Intraband processes}
 $\bullet$ When each electron stays in its band, $n'=n$, the  Bloch-state scalar product $\langle u_{n',\textbf{k}'}|u_{n,\textbf{k}} \rangle$ is close to 1; so,
 \begin{equation}\label{app:4}
 \mathcal{V}(n,\textbf{k}+\textbf{q};n,\textbf{k})    \simeq 
 \sqrt{\frac{4\pi e^2}{L^3 q^2}}
 \end{equation}
 As a result, the \textit{intra}band Coulomb scatterings reduce to $4\pi e^2/L^3 q^2$, a result incorrectly thought as obvious.

\begin{figure}
\includegraphics[width=.35\textwidth]{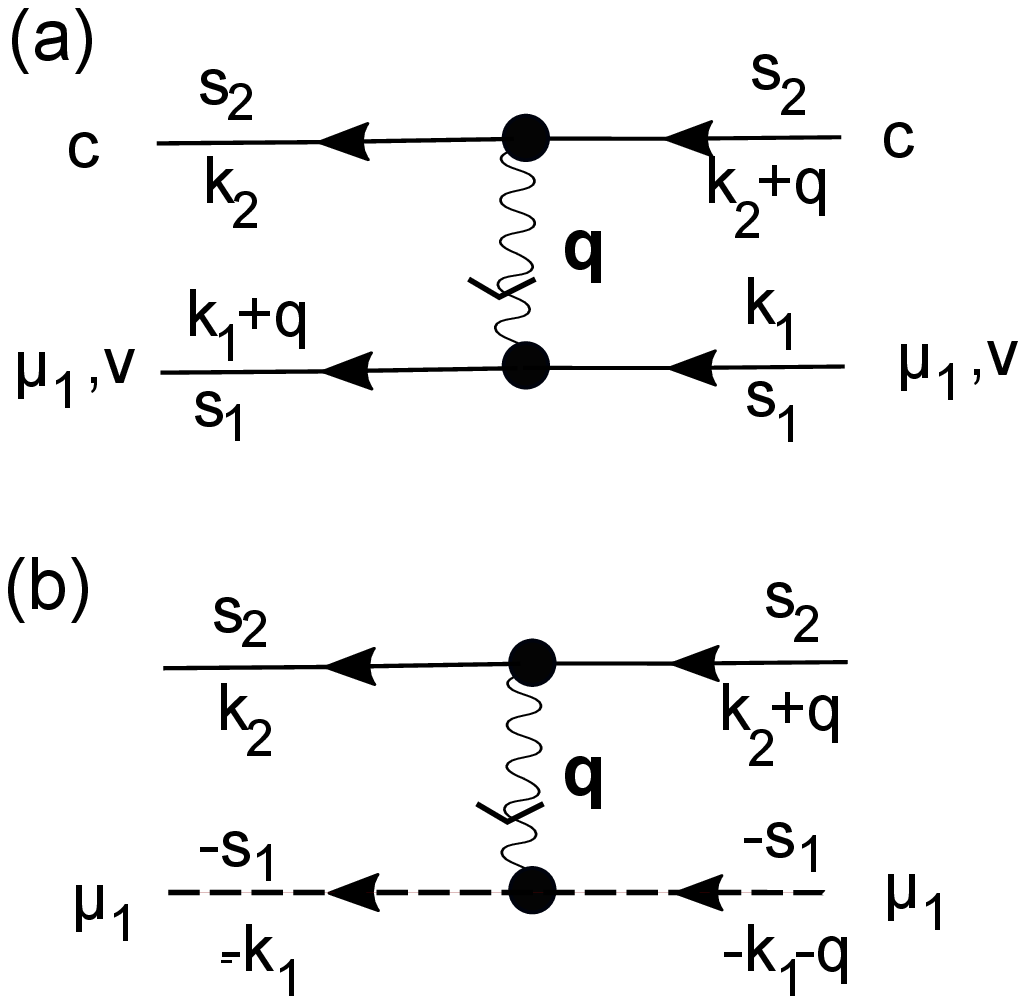}
\caption{Diagrammatic representation of the intraband Coulomb interaction (a) between conduction and valence electrons, as given in Eq.~(\ref{app:5}) and (b) between electron and hole, according to Eq.~(\ref{app:12}). Solid lines represent electrons and dashed lines represent holes. Intraband Coulomb processes exist whatever the electron spin $s_2$ and the hole spin $-s_1$.}
\label{fig:S1}
\end{figure}

$\bullet$ By noting that in Eq.~(\ref{app:2}), we can have ($n_2=c$, $n_1=(v,\mu)$) or ($n_2=(v,\mu)$, $n_1=c$), the \textit{intra}band Coulomb interaction between conduction and valence electrons reduces, in the small \textbf{q} limit, to
\begin{eqnarray}
\label{app:5}
\hat{V}^{(intra)}_{Coul} \simeq \sum_{\textbf{q}\not=0} \frac{4\pi e^2}{L^3q^2} \sum_{\mu_1,\textbf{k}_1,s_1} \,\,\sum_{\textbf{k}_2,s_2}
\,\,\,\,\,\, \,\,\,\,\,\, \,\,\,\,\,\, \,\,\,\,\,\, \,\,\,\,\,\, \,\,\,\,\,\, \,\,\,\,\,\, 
\\
\hat{a}^\dag_{v,\textbf{k}_1+\textbf{q};\mu_1,s_1} \hat{a}^\dag_{c,\textbf{k}_2;s_2}
 \hat{a}_{c,\textbf{k}_2+\textbf{q};s_2}  \hat{a}_{ v,\textbf{k}_1;\mu_1,s_1}
 \nonumber
 \end{eqnarray}
this scattering being diagonal in the valence electron spatial  index $\mu$. Its diagrammatic representation is shown in Fig.~S\ref{fig:S1}(a).

\subsection{Interband processes}
$\bullet$ When the electrons change band, $n' \neq n$, each Bloch-state scalar product scales as $q$; so, the \textit{inter}band Coulomb scattering stays finite in the small $\textbf{q}$ limit:
\begin{equation}
\mathcal{V}(c,\textbf{k}+\textbf{q}; v,\textbf{k},\mu)  \simeq    \frac{\hbar\textbf{q}}{m_0}\cdot
\frac{\langle u_{c,\textbf{k}}|\hat{\textbf{p}}| u_{ v,\textbf{k};\mu}\rangle }{\varepsilon_{c,\textbf{k}}-\varepsilon_{v,\textbf{k}}}\sqrt{\frac{4\pi e^2}{L^3 q^2}}
\end{equation}
 The energy difference $(\varepsilon_{c,\textbf{k}}-\varepsilon_{v,\textbf{k}})$ is close to the band gap $E_{gap}$. The  matrix element of the momentum operator $\hat{\textbf{p}}=-i\hbar\nabla$ between conduction and valence states, is obtained by writing $\hat{\textbf{p}}$ as $\sum_{\mu ' }  \hat{p}_{\mu '} \textbf{e}_{\mu '}$ with
 \begin{equation}
 \label{app:7}
 \langle u_{c,\textbf{0}} | \hat{p}_{\mu '} | u_{v,\textbf{0};\mu} \rangle = \delta_{\mu',\mu } P_{cv}
 \end{equation}
$P_{cv}$ being $\mu$ independent for cubic crystals. This gives 
\begin{eqnarray}
\mathcal{V}(c,\textbf{k}+\textbf{q};v,\textbf{k},\mu )
\simeq
 \frac{\Lambda}{E_{gap}} \,\, \left( \textbf{e}_{\mu }{\cdot} \frac{\textbf{q}}{q}\right) 
 \end{eqnarray}
with $\Lambda$ equal to 
 \begin{equation}
 \Lambda = \frac{\hbar P_{cv}}{m_0}\sqrt{\frac{4\pi e^2}{L^3}}
 \end{equation}

In the following, it will be useful to note that in MKS units instead of CGS units, that is, for $e^2/r$ written as $q_e^2/(4\pi\epsilon_0 r)$, the $\Lambda$ prefactor reads as 
\begin{equation}
\label{app:8}
\Lambda = \frac{q_e P_{cv}}{m_0}\sqrt{\frac{\hbar^2}{\epsilon_0L^3}}
\end{equation}

\begin{figure}
\includegraphics[width=0.35\textwidth]{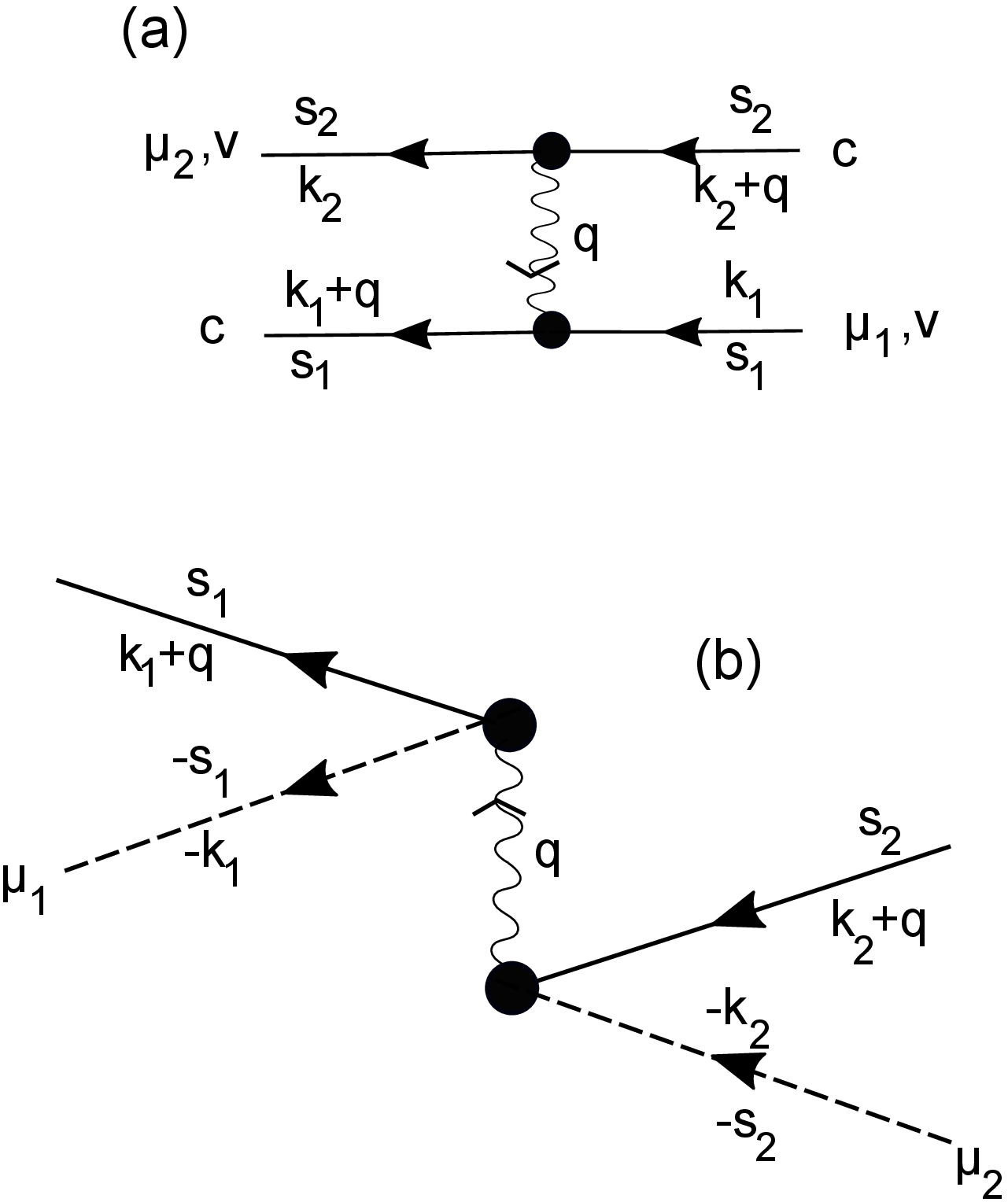}
\caption{Diagrammatic representation of the \textit{inter}band Coulomb interaction (a) between conduction and valence electrons, as given in Eq.~(\ref{app:11}) and (b) between electron and hole as given in Eq.~(\ref{app:13}). The interband Coulomb interaction only acts on electron-hole pairs in a spin-singlet state.}
\label{fig:S2}
\end{figure}

$\bullet$ By again noting that in Eq.~(\ref{app:2}), the indices ($n_2,n_1$) can be equal to $(c,v)$ or $(v,c)$ while $P_{v,c}=P_{c,v}^\ast$, the above equations yield the interband Coulomb interaction as
\begin{eqnarray}
\label{app:11}
\hat{V}_{Coul}^{(inter)} = \frac{|\Lambda|^2}{E_{gap}^2} \sum_\textbf{q} \sum_{\mu_1,\mu_2} \left( \frac{\textbf{q}}{q}\cdot\textbf{e}_{\mu_1}\right)\left( \frac{\textbf{q}}{q}\cdot\textbf{e}_{\mu_2}\right)\nonumber\\
\sum_{\textbf{k}_1,s_1} \sum_{\textbf{k}_2,s_2} \hat{a}^\dag_{c,\textbf{k}_1+\textbf{q};s_1} \hat{a}^\dag_{ v,\textbf{k}_2;\mu_2,s_2}
 \hat{a}_{c,\textbf{k}_2+\textbf{q};s_2}  \hat{a}_{ v,\textbf{k}_1;\mu_1,s_1} 
\end{eqnarray}
Its diagrammatic representation is shown in Fig.~S\ref{fig:S2}(a). This scattering depends on the wave-vector transfer $\textbf{q}$ and the hole index $\mu$ as $(\textbf{q}/{q})\cdot \textbf{e}_{\mu}$. So, unlike intraband Coulomb scattering, the interband scattering remains finite in the small \textbf{q} limit. 

$\bullet$ The next step is to turn from conduction and valence electrons to electrons and holes. This is done according to
\begin{equation}
\label{app:12}
\hat{a}^\dag_{c,\textbf{k};s} = \hat{a}^\dag_{\textbf{k},s} \hspace{.9cm} \hat{a}_{ v,\textbf{k};\mu,s} = (-1)^{1/2-s} \hat{b}^\dag_{-\textbf{k},\mu,-s}
\end{equation} 
The \textit{inter}band Coulomb interaction between electrons and holes then reads as
\begin{eqnarray}
\label{app:13}
\hat{V}_{Coul}^{(inter)} = \frac{|\Lambda|^2}{E_{gap}^2} \sum_\textbf{q} \sum_{\mu_1,\mu_2} \left( \frac{\textbf{q}}{q}\cdot\textbf{e}_{\mu_1}\right)\left( \frac{\textbf{q}}{q}\cdot\textbf{e}_{\mu_2}\right)\nonumber\\
\sum_{\textbf{k}_1,s_1} (-1)^{1/2-s_1} \hat{a}^\dag_{\textbf{k}_1+\textbf{q},s_1} \hat{b}^\dag_{-\textbf{k}_1,\mu_1,-s_1}\nonumber\\
\sum_{\textbf{k}_2,s_2} (-1)^{1/2-s_2}  \hat{b}_{-\textbf{k}_2,\mu_2,-s_2}\hat{a}_{\textbf{k}_2+\textbf{q},s_2}
\end{eqnarray}
This interaction, shown in Fig.~S\ref{fig:S2}(b), scatters an electron-hole pair having a $\mu_2$ hole and a center-of-mass wave vector \textbf{K} equal to \textbf{q} into another pair having a $\mu_1$ hole and the same center-of-mass wave vector.

It is worth noting that the electron-hole pairs involved in the \textit{inter}band interaction are in a spin-singlet state $(S=0,S_z=0)$
\begin{eqnarray} 
 \hat{B}^\dag_{S=0}=\frac{\hat{a}^\dag_{1/2}\hat{b}^\dag_{-1/2}-\hat{a}^\dag_{-1/2}\hat{b}^\dag_{1/2}}{  \sqrt{2}  }
\end{eqnarray}

$\bullet$ To go further, we introduce the creation operator for an electron-hole pair made of a $\mu$ hole, with pair center-of-mass wave vector \textbf{K} and relative-motion wave vector \textbf{k}, that reads in terms of electron and hole operators as 
\begin{equation}
\label{app:15}
 \hat{a}^\dag_{\textbf{k}+\gamma_e \textbf{K}}b^\dag_{-\textbf{k}+\gamma_h \textbf{K},\mu} \equiv \hat{B}^\dag_{\textbf{K},\textbf{k};\mu}
\end{equation}
with $\gamma_e=1-\gamma_h=m_e/(m_e+m_h)$ for electron and hole masses ($m_e$,$m_h$) in order to separate the center-of-mass and relative-motion kinetic energies as
\begin{equation}
\frac{\hbar^2(\textbf{k}+\gamma_e \textbf{K})^2}{2m_e} + \frac{\hbar^2(-\textbf{k}+\gamma_h \textbf{K})^2}{2m_h} = \frac{\hbar^2\textbf{K}^2}{2M_{eh}} + \frac{\hbar^2\textbf{k}^2}{2\mu_{eh}}
\end{equation}
with $M_{eh}$=($m_e+m_h$) and $\mu_{eh}^{-1}$=$m_e^{-1}$+$m_h^{-1}$. The electron-hole pair operator in Eq.~(\ref{app:13}) then reads 
\begin{equation}
\label{app:17}
\sum_{\textbf{k},s} (-1)^{1/2-s} \hat{a}^\dag_{\textbf{k}+\textbf{K},s} \hat{b}^\dag_{-\textbf{k},\mu,-s}
= \sqrt{2}\sum_{\textbf{k}'} \hat{B}^\dag _{\textbf{K},\textbf{k}';\mu,S=0}
\end{equation}

This gives the interband Coulomb interaction in terms of singlet electron-hole pairs as
\begin{eqnarray}
\label{app:18}
\hat{V}_{Coul}^{(inter)} = 2\,\frac{|\Lambda|^2}{E_{gap}^2} \sum_\textbf{K} \sum_{\mu_1,\mu_2} \left( \frac{\textbf{K}}{K}\cdot\textbf{e}_{\mu_1}\right)\left( \frac{\textbf{K}}{K}\cdot\textbf{e}_{\mu_2}\right)\nonumber\\
\sum_{\textbf{k}_1} \hat{B}^\dag _{\textbf{K},\textbf{k}_1;\mu_1,S=0}
\,\,
\sum_{\textbf{k}_2}\hat{B} _{\textbf{K},\textbf{k}_2;\mu_2,S=0}
\,\,\,\,
\end{eqnarray}
This equation evidences that the interband Coulomb interaction affects spin-singlet pairs  only, with a probability that depends on the hole spatial state $\mu$ through $\textbf{e}_{\mu} \cdot (\textbf{K}/K)$ for a pair center-of-mass wave vector \textbf{K}.

\section{Photon-electron coupling}
\subsection{General form}
$\bullet$ The classical form of the electron-photon interaction, at first order in the transverse electromagnetic field, is given in the Coulomb gauge by
\begin{equation}
W_t= -\frac{q_e}{m_0}\sum_{j} \textbf{A}_\bot(\textbf{r}_{j}(t),t)\cdot \textbf{p}_{j}(t)
\end{equation}
for electrons located at $\textbf{r}_{j}(t)$ with free mass $m_0$, charge $q_e$ and momentum $\textbf{p}_{j}(t)$. 

The $ \textbf{A}(\textbf{r},t)$ vector potential reads in second quantization as
 \begin{equation}
\hat{ \textbf{A}}(\textbf{r})=\sum_\textbf{Q} \mathcal{A}_\textbf{Q} \sum_{\lambda=(X,Y)}
 \textbf{e}_{\lambda,\textbf{Q}}
  \Big( \hat{\alpha}_{_{\textbf{Q},\lambda}}
 e^{i\textbf{Q}\cdot\textbf{r}}
 + \hat{\alpha}^\dag_{_{\textbf{Q},\lambda}} e^{-i\textbf{Q}\cdot\textbf{r}} \Big) 
 \end{equation}
 with $\mathcal{A}_\textbf{Q}=\sqrt{\hbar/2\epsilon_0\omega_\textbf{Q} L^3}$ for photon with energy $\hbar \omega_\textbf{Q}$ in a sample volume $L^3$. The operator 
 $ \hat{\alpha}^\dag_{_{\textbf{Q},\lambda}}$ creates a  photon with wave vector $\textbf{Q}$ in one of the two polarizations $ \textbf{e}_{\lambda,\textbf{Q}}$ perpendicular to $ \textbf{e}_{Z,\textbf{Q}}=\textbf{Q}/Q$. 
 
$\bullet$  This gives the photon-electron coupling $W_t$ in second quantization as
\begin{equation}
\hat{W}_{ph}=-\frac{q_e}{m_0}\sum_\textbf{Q}\mathcal{A}_\textbf{Q}\sum_{\lambda=(X,Y)}
\Big(\hat{\alpha}_{_{\textbf{Q},\lambda}} \, \textbf{e}_{\lambda,\textbf{Q}} {\cdot}\hat{\textbf{P}}_\textbf{Q} +h.c\Big)
\end{equation}
where $\hat{\textbf{P}}_\textbf{Q}$=$\sum_{j} e^{i\textbf{Q}\cdot\textbf{r}_j}\hat{\textbf{p}}_j$ is a one-body operator that does not act on spin. In second quantization, it reads in terms of creation operators $\hat{a}_{n,\textbf{k},s}^\dag$ for electrons with spin $s$ in the Bloch state $|n,\textbf{k} \rangle$  as
\begin{equation}
\hat{\textbf{P}}_{\textbf{Q}}=\sum_s
\sum_{n',n}\,\,
\sum_{\textbf{k}',\textbf{k}}
\langle n',\textbf{k}'| e^{i\textbf{Q}\cdot\textbf{r}} \hat{\textbf{p}}|n,\textbf{k}\rangle
 \hat{a}^\dag_{n',\textbf{k}',s}\hat{a}_{n,\textbf{k},s}
\end{equation}
 the prefactor being given, using Eq.~(\ref{1}), by
\begin{equation}
\langle n',\textbf{k}'| e^{i\textbf{Q}\cdot\textbf{r}} \hat{\textbf{p}}|n,\textbf{k}\rangle
 \simeq \delta_{\textbf{k}', \textbf{k}+\textbf{Q}}
\langle u_{n',\textbf{k}'}|\hbar\textbf{k}+\hat{\textbf{p}}|u_{n,\textbf{k}} \rangle
\end{equation}

\subsection{Interband coupling}

$\bullet$ Let us concentrate on the \textit{inter}band coupling, that is, $(n', n)$ equal to $(c,v)$ or  $(v,c)$. For threefold valence states characterized by $\mu$=$(x,y,z)$, the matrix element $\langle u_{c,\textbf{k}+\textbf{Q}}|\hbar\textbf{k}+\hat{\textbf{p}}|u_{ v,\textbf{k};\mu} \rangle$ reduces for (\textbf{k},\textbf{Q}) small, to $\langle u_{c,\textbf{0}}|\hat{\textbf{p}} |u_{v,\textbf{0};\mu} \rangle$, that is, according to Eq.~(\ref{app:7}), 
\begin{equation}  \label{app:24}
\langle u_{c,\textbf{0}}|\hat{\textbf{p}}|u_{v,\textbf{0};\mu} \rangle
=\langle u_{c,\textbf{0}}|\sum_{\mu' }\textbf{e}_{\mu' }\hat{p}_{\mu' } |u_{ v,\textbf{0};\mu} \rangle
=\textbf{e}_{\mu }P_{c,v}
\end{equation}

The above equations give the interband part of the electron-photon interaction as 
\begin{align}
\hat{W}_{ph}^{(inter)}{=}\frac{-q_e}{m_0}\sum_\textbf{Q}\mathcal{A}_\textbf{Q}\sum_{\lambda{=}(X,Y)} \hat{\alpha}_{_{\textbf{Q},\lambda}} 
\sum_{\mu=(x,y,z)} (\textbf{e}_{\lambda,\textbf{Q} } \cdot \textbf{e}_{\mu })
\nonumber \\
\sum_{\textbf{k},s}\Big(P_{c,v}\hat{a}^\dag_{c,\textbf{k}{+}\textbf{Q};s}\hat{a}_{ v,\textbf{k};\mu,s} {+}P_{v,c}\hat{a}^\dag_{ v,\textbf{k}+\textbf{Q};\mu,s}\hat{a}_{c,\textbf{k};s}\Big)
 +h.c  \,\,\,\,\,\,\,\,\,
 \end{align}
The absorption of a photon goes along with not only the excitation of a valence electron (term in $P_{c,v}$ called ``resonant''), but also the de-excitation of a conduction electron (term in $P_{v,c}$ called ``nonresonant''). 
This leads us to split $\hat{W}_{ph}^{(inter)}$ into resonant and nonresonant terms, 
\begin{equation}
\hat{W}^{(inter)}_{ph}=\hat{W}^{(inter)}_{res}+\hat{W}^{(inter)}_{nres}
\end{equation}

Before going further, we  note that 
\begin{equation}
\frac{q_e}{m_0}\mathcal{A}_\textbf{Q}P_{c,v}
=  \frac{q_e}{m_0}    \sqrt{\frac{\hbar}{2\epsilon_0 
\omega_\textbf{Q} 
L^3}} \,\, 
P_{c,v}
=\frac{\Lambda}{ \sqrt{2\hbar \omega_\textbf{Q}}     }
\end{equation} 
where $\Lambda$ is the quantity appearing in the interband Coulomb interaction, but written in MKS units (see Eq.~(\ref{app:8})). 

\begin{figure}
\includegraphics[width=.48\textwidth]{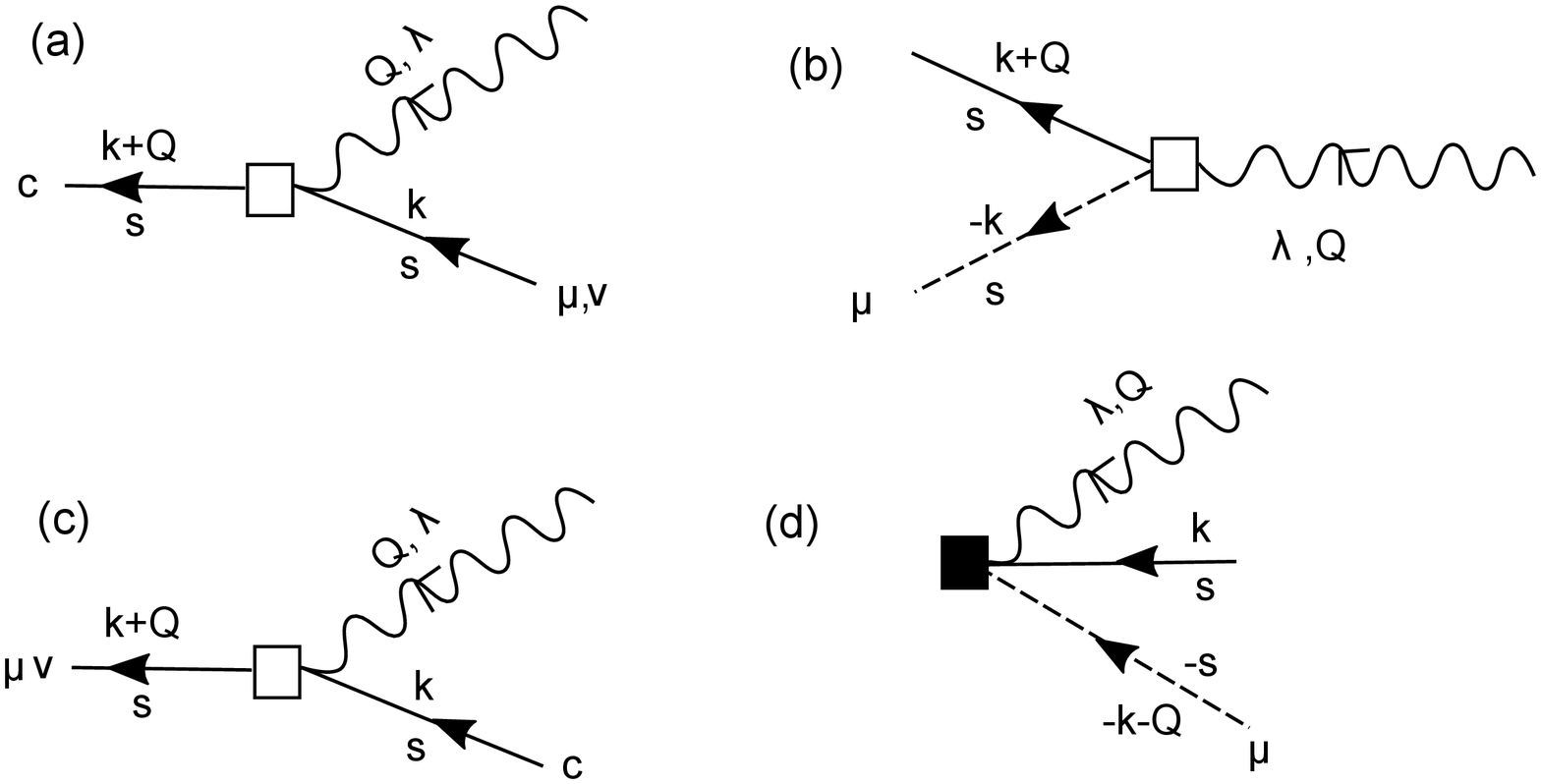}
\caption{Diagrammatic representation of the photon-electron coupling. Resonant absorption of a photon with wave vector \textbf{Q} and polarization vector $\textbf{e}_{\lambda,\textbf{Q}}$, (a) in terms of conduction and valence electrons, as given in Eq.~(\ref{app:28}) and (b) in terms of electron and hole, as given in Eq.~(\ref{app:30}). (c-d) Similar diagrams for nonresonant absorption.
}\label{fig:S3}
\end{figure}

$\bullet$ The  resonant part of the interband electron-photon interaction appears as
\begin{eqnarray}
\label{app:28}
\hat{W}^{(inter)}_{res} &=& -\Lambda \sum_\textbf{Q} \sum_{\lambda=(X,Y)} \sum_{\mu=(x,y,z)} \frac{\textbf{e}_{\lambda,\textbf{Q}}\cdot\textbf{e}_{\mu}}{\sqrt{2\hbar\omega_\textbf{Q}}}
\nonumber\\
&{}&
\hat{\alpha}_{_{\textbf{Q},\lambda}}
\sum_{\textbf{k},s} \hat{a}^\dag_{c,\textbf{k}+\textbf{Q};s}\hat{a}_{v,\textbf{k};\mu,s} +h.c\nonumber\\
&\equiv& \hat{U}_{res}+\hat{U}^\dag_{res}\,\,\,\,\,\,\,\,\,\,\,\,\,\,\,\,\,\,\,\,\,\,\,\,\,\,\,\,\,\,\,\,\,\,\,\,\,\,\,\,\,\,\,\,\,\,\,\,\,\,\,\,\,\,\,\,\,\,\,\,\,\,\,\,\,\,\,\,\,\,\,\,
\end{eqnarray}
The operator $\hat{U}_{res}$ absorbs a photon and excites an electron from the valence band to the conduction band (see Fig.~S\ref{fig:S3}(a)).

In the same way, the nonresonant part of the interband electron-photon interaction reads as

\begin{eqnarray}
\label{app:29}
\hat{W}^{(inter)}_{nres} = -\Lambda^\ast \sum_\textbf{Q} \sum_{\lambda=(X,Y)} \sum_{\mu=(x,y,z)} \frac{\textbf{e}_{\lambda,\textbf{Q}}\cdot\textbf{e}_{\mu}}{\sqrt{2\hbar\omega_\textbf{Q}}}
\nonumber\\
\hat{\alpha}_{_{\textbf{Q},\lambda}}
\sum_{\textbf{k},s}
\hat{a}^\dag_{v,\textbf{k}+\textbf{Q};\mu ,s}\hat{a}_{c,\textbf{k};s}
+h.c\nonumber\\
\equiv \hat{U}_{nres}+\hat{U}^\dag_{nres}\,\,\,\,\,\,\,\,\,\,\,\,\,\,\,\,\,\,\,\,\,\,\,\,\,\,\,\,\,\,\,\,\,\,\,\,\,\,\,\,\,\,\,\,\,\,\,\,\,\,\,\,\,\,\,\,\,\,\,\,\,\,\,\,
\end{eqnarray}
The operator $\hat{U}_{nres}$ absorbs a photon and de-excites an electron from the conduction band to the valence band (see Fig.~S\ref{fig:S3}(c)).

$\bullet$ By turning from conduction and valence electrons to electrons and holes, the absorption of a photon with wave vector \textbf{Q} in resonant processes goes along with the creation of an electron-hole pair with center-of-mass wave vector \textbf{Q}, namely
\begin{eqnarray}
\label{app:30}
\hat{U}_{res}= -\Lambda \sum_\textbf{Q} \sum_{\lambda=(X,Y)} \sum_{\mu=(x,y,z)} \frac{\textbf{e}_{\lambda,\textbf{Q}}\cdot\textbf{e}_{\mu}}{\sqrt{\hbar\omega_\textbf{Q}}}
\hspace{1cm}\\
\hat{\alpha}_{_{\textbf{Q},\lambda}}
\,\,
\frac{1}{\sqrt{2}  }
\sum_{\textbf{k},s} (-1)^{1/2-s}
 \hat{a}^\dag_{\textbf{k}+\textbf{Q},s}
  \hat{b}^\dag_{-\textbf{k},\mu,-s} \nonumber
\end{eqnarray}
(see Fig.~S\ref{fig:S3}(b)). Since the photon is spinless, the created pair also is in a singlet state (see Eq.~(\ref{app:17})); so, the above equation can be rewritten as
\begin{eqnarray}
\label{app:31}
\hat{U}_{res}= -\Lambda \sum_\textbf{Q} \sum_{\lambda=(X,Y)} \sum_{\mu=(x,y,z)} \frac{\textbf{e}_{\lambda,\textbf{Q}}\cdot\textbf{e}_{\mu}}{\sqrt{\hbar\omega_\textbf{Q}}}
\nonumber\\
\hat{\alpha}_{_{\textbf{Q},\lambda}}
\sum_{\textbf{k}} \hat{B}^\dag_{\textbf{Q},\textbf{k};\mu,S=0}
\end{eqnarray}

$\bullet$ In the case of nonresonant processes, the photon absorption goes along with the \textit{recombination} of an electron-hole pair also in a singlet state, its center-of-mass wave vector being equal to ($-\textbf{Q}$), in order to conserve the photon-pair wave vector (see Fig.~S\ref{fig:S3}(d)),
\begin{eqnarray}
\hat{U}_{nres} = -\Lambda^\ast \sum_\textbf{Q} \sum_{\lambda=(X,Y)} \sum_{\mu=(x,y,z)} \frac{\textbf{e}_{\lambda,\textbf{Q}}\cdot\textbf{e}_{\mu}}{\sqrt{\hbar\omega_\textbf{Q}}}\nonumber
\\
\hat{\alpha}_{_{\textbf{Q},\lambda}}\,\sum_{\textbf{k}} \hat{B}_{-\textbf{Q},\textbf{k};\mu,S=0}\label{app:32}
\end{eqnarray}

\section{Excitons}

\begin{figure}
\includegraphics[width=.48\textwidth]{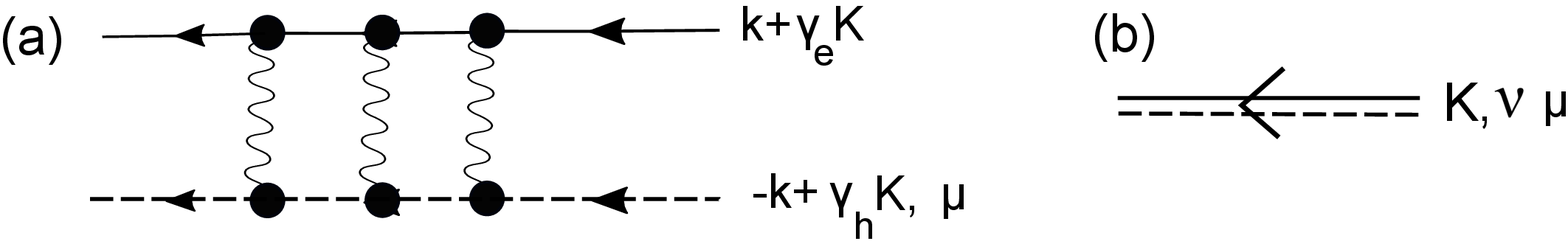}
\caption{(a) Repeated intraband Coulomb processes transform one free electron-hole pair with center-of-mass wave vector  \textbf{K}, relative-motion wave vector \textbf{k} and hole orbital index $\mu$, into an exciton (b) with the same characteristics, its relative-motion index being $\nu$ (see Eqs.~(\ref{app:15},\ref{app:33})).}\label{fig:S4}
\end{figure}

$\bullet$ The repeated intraband Coulomb interactions transform a free electron-hole pair into a correlated pair, that is, an exciton (see Fig.~S\ref{fig:S4}). Since Coulomb interaction conserves wave vector and spin, the free and correlated pairs have the same center-of-mass wave vector and spin. Let us here concentrate on the spin-singlet states because these are the ones involved in the interband processes.

The creation operators for  pairs with relative-motion wave vector $\textbf{k}$ and pairs correlated by intraband processes into excitons with relative-motion index $\nu$, are related by 
\begin{align}
\label{app:33}
\hat{B}^\dag_{\textbf{K},\textbf{k};\mu,S=0} &= \sum_\nu \hat{B}^\dag_{\textbf{K},\nu;\mu,S=0} \langle \nu | \textbf{k} \rangle\\
\hat{B}^\dag_{\textbf{K},\nu;\mu,S=0}& = \sum_\textbf{k} \hat{B}^\dag_{\textbf{K},\textbf{k};\mu,S=0} \langle \textbf{k} | \nu \rangle
\end{align}
where the $ \langle \textbf{k}|\nu\rangle$ wave function in momentum space fulfills
\begin{equation}
0= \left( \frac{\hbar^2\textbf{k}^2}{2\mu_{eh}}-\varepsilon_\nu \right) \langle \textbf{k} | \nu \rangle - \sum_{\textbf{k}'\neq \textbf{k}} \frac{4\pi e^2}{L^3|\textbf{k}'-\textbf{k}|^2} \langle \textbf{k}' | \nu \rangle
\end{equation}
the Coulomb scattering being further reduced by the semiconductor dielectric constant $\epsilon_{sc}$ that comes from bubble-like interband Coulomb processes \cite{appCombescot_book}. 

The energy of the spin-singlet exciton
\begin{equation}
| X_{\textbf{K},\nu;\mu,S=0}\rangle = \hat{B}^\dag_{\textbf{K},\nu;\mu,S=0}|v\rangle
\end{equation}
where $|v\rangle$ denotes the vacuum state, is equal to 
\begin{equation}
E_{gap}+ \frac{\hbar^2\textbf{K}^2}{2M_{eh}}+\varepsilon_\nu
\equiv  E_{\textbf{K},\nu}
\end{equation}
 with $\varepsilon_\nu$ negative for bound states; the exciton energy does not depend on $\mu$ for cubic crystals.

$\bullet$ By using Eq.~(\ref{app:33}) in the various interband operators, we can replace their electron-hole pairs by
\begin{equation}
\label{app:37}
\Lambda^\ast  \sum_{\textbf{k}} \hat{B}_{\textbf{Q},\textbf{k};\mu,S=0} = \sum_{\nu} \Lambda^\ast_\nu  \hat{B}_{\textbf{Q},\nu;\mu,S=0}
\end{equation}
with $\Lambda^\ast_{\nu}$ given by
\begin{equation}
\Lambda^\ast_{\nu}=  \Lambda^\ast \sum_\textbf{k}\langle \textbf{k}| \nu \rangle=\Lambda^\ast L^{3/2} \langle \textbf{r}=\textbf{0}| \nu \rangle
\end{equation}
as obtained from $\langle \textbf{r} |\textbf{k} \rangle=e^{i\textbf{k}\cdot \textbf{r}}/L^{3/2}$; so, the sum over $\textbf{k}$ also reads $L^{3/2}$ $\sum_\textbf{k} \langle \textbf{r}=\textbf{0} | \textbf{k} \rangle\langle \textbf{k} | \nu \rangle=L^{3/2}\langle \textbf{r}=\textbf{0} | \nu \rangle$.

When used in the $\hat{U}_{res}$ operator given in Eq.~(\ref{app:31}), we find that it also reads
\begin{eqnarray}
\hat{U}_{res} 
= - \sum_\textbf{Q} \sum_{\lambda=(X,Y)} \sum_{\mu=(x,y,z)} \textbf{e}_{\lambda,\textbf{Q}}\cdot\textbf{e}_{\mu}\,\,\,\,\,\,\,
\,\,\,\,\,\,\,\,\,\,\,\,\,\,\,\,\,\,\,\,\,
\nonumber\\
\hat{\alpha}_{_{\textbf{Q},\lambda}}
 \sum_{\nu} \hat{B}^\dag_{\textbf{Q},\nu;\mu,S=0} 
 \frac{\Lambda_{\nu}}{\sqrt{\hbar\omega_\textbf{Q}}}
  \,\,\,\,\,\,\,
\end{eqnarray}

In the following, it will be of interest to note that the exciton-photon Rabi coupling between a \textit{photocreated} ground-state exciton $\nu_0$ in a  spin-singlet state, and an \textit{absorbed} photon with energy $\hbar\omega_\textbf{Q}$ close to the band gap, is related to $\Lambda_{\nu}$ through
\begin{equation}
\label{app:41}
\Omega_{ph-X}
\propto
\frac{\Lambda_{\nu_0}}{\sqrt{E_{gap}}}
\end{equation}
within a factor that depends on the number of photons.

\section{Energy shift of bright excitons}

We now have all the necessary knowledge to derive the energy change induced by the interaction between the electromagnetic field and a ground-state exciton in the spin-singlet state, that is, in the subspace made of $| X_{\textbf{K},\nu_0;\mu,S=0}\rangle$ excitons, with energy $E_{\textbf{K},\nu_0}$ whatever $\mu$.

\subsection { Interband Coulomb contribution}

The interband Coulomb interaction, given in Eq.~(\ref{app:18}) destroys a singlet electron-hole pair  with a $\mu$ hole and creates a similar pair with a $\mu'$ hole. Using Eq.~(\ref{app:37}), we then find
\begin{eqnarray}
\label{app:42}
 \Delta_{Coul }^{(\mu',\mu)} &\equiv&\langle X_{\textbf{K},\nu_0;\mu',S=0}| \hat{V}_{Coul}^{(inter)} | X_{\textbf{K},\nu_0;\mu,S=0} \rangle 
\nonumber 
\\
&=&2\,\frac{|\Lambda_{\nu_0}^2|}{E_{gap}^2}\left( \frac{\textbf{K}}{K}\cdot \textbf{e}_{\mu'} \right)
\left( \frac{\textbf{K}}{K}\cdot \textbf{e}_\mu \right)
\end{eqnarray}

\subsection{Interband photon-electron contribution}

$\bullet$ In the interband electron-photon interaction $\hat{W}^{(inter)}_{ph}$, the creation of a photon goes along with resonant and nonresonant processes, in which an electron-hole pair is either destroyed or created. Thus, the change in energy induced by the photon coupling to excitons cancels at first order in the coupling. 

At second order, the change reads
\begin{eqnarray}\label{app:43}
\Delta_{ph}^{(\mu',\mu )} =
\,\,\,\,\,\,\,\,\,\,\,\,\,\,\,\,\,\,\,\,\,\,\,\,\,\,\,\,\,\,\,\,\,\,\,\,\,\,\,\,\,\,\,\,\,\,\,\,\,\,\,\,\,\,\,\,\,\,\,\,\,\,\,\,\,\,\,\,\,\,\,\,\,\,\,\,\,\,\,\,\,\,\,\,\,\,\,\,\,\,\,\,\,\,\,\,\,\,\,\,\,\,\,\,\,\,\,\,\,\,\,\,
\\
 \sum_f \frac{\langle X_{\textbf{K},\nu_0;\mu',S=0} | \hat{W}^{(inter)}_{ph} | f \rangle \langle f | \hat{W}^{(inter)}_{ph} | X_{\textbf{K},\nu_0;\mu,S=0} \rangle}{E_{\textbf{K},\nu_0}-E_f}
 \nonumber
\end{eqnarray}
with $\hat{W}_{ph}^{(inter)}$ replaced by $ \hat{W}^{(inter)}_{res}$ or  $ \hat{W}^{(inter)}_{nres}$ on both sides, in order to keep the number of electron-hole pairs.

 $\bullet$ \textit{Resonant processes}

\begin{figure}
\includegraphics[width=.5\textwidth]{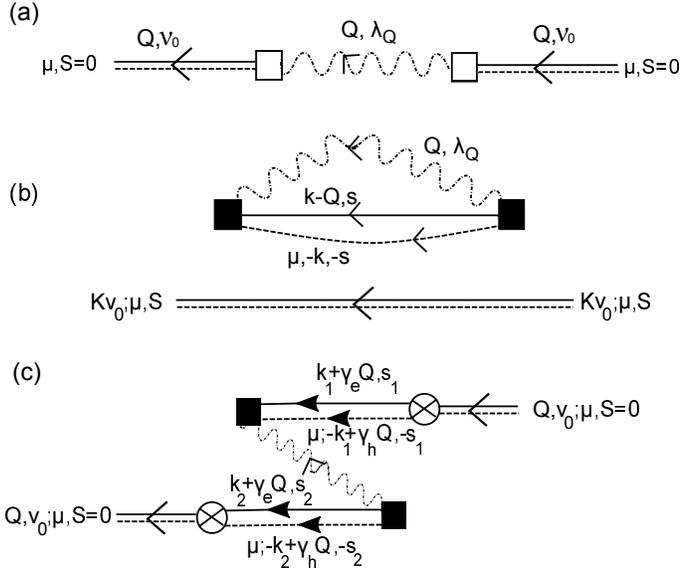}
\caption{(a) Resonant coupling to a virtual photon with wave vector $\textbf{Q}$ that produces a shift to singlet exciton ($S=0$). Nonresonant coupling to a virtual photon: direct process (b) and indirect process (c). In (b), the exciton does not interact with the photon, as seen from the fact that the photon and the exciton have different wave vectors, $\textbf{Q}$ and $\textbf{K}$; so, direct nonresonant processes do not produce a shift, while in (c), they do interact; so, the incoming and outgoing excitons have a wave vector $\textbf{Q}$, while the virtual photon has a wave vector $-\textbf{Q}$. }\label{fig:S5}
\end{figure}

From Eqs.~(\ref{app:28},\ref{app:31},\ref{app:37}), we get
\begin{eqnarray}\label{44}
\hat{W}^{(inter)}_{res}|X_{\textbf{K},\nu_0;\mu,S=0}\rangle=
\hat{U}^\dag_{res}|X_{\textbf{K},\nu_0;\mu,S=0}\rangle  
\,\,\,\,\,\,\,\,\,\,\,\,\\
=-  \frac{\Lambda^*_ {\nu_0}}{\sqrt{\hbar\omega_{\textbf{K}}}} 
\sum_{\lambda=(X,Y)}
 (\textbf{e}_{\lambda,{\textbf{K}}}\cdot\textbf{e}_{{\mu}})
 \,\,
 \hat{\alpha}^\dag_{_{\textbf{K},\lambda}}
 |v\rangle\nonumber
\end{eqnarray}
So, the intermediate state $|f\rangle= \hat{\alpha}^\dag_{_{\textbf{K},\lambda}}
 |v\rangle$ corresponds to a \textbf{K} photon with energy $\hbar\omega_\textbf{K}$ (see Fig.~S\ref{fig:S5}(a)).

By using a similar result for $\langle X_{\textbf{K},\nu_0;\mu',S=0} | \hat{U}_{res}$, which imposes the two photons to have the same polarization, we end with resonant electron-photon processes producing a shift given by
\begin{eqnarray}\label{app:45}
\Delta_{res}^{(\mu',\mu)} = \frac{|\Lambda_{\nu_0}|^2}{\hbar\omega_\textbf{K}}\frac{1}{E_{\textbf{K},\nu_0}-\hbar\omega_\textbf{K}}
\,\,\,\,\,\,\,\,\,\,\,\,\,\,\,\,\,\,\,\,\,\,\,\,\,\,\,\,\,\,\,\,\,\,\,\,\,\,\,\,\,\,\,\,\,
\\
 \sum_{\lambda=(X,Y)} \left(  \textbf{e}_{\lambda,\textbf{K}}\cdot\textbf{e}_  {\mu'} \right)
\left(  \textbf{e}_{\lambda,\textbf{K}}\cdot\textbf{e}_\mu \right)
\nonumber
\end{eqnarray}

$\bullet$\textit{ Nonresonant processes}

In the same way, we find from Eqs.~(\ref{app:29},\ref{app:32},\ref{app:37}) that the nonresonant electron-photon coupling leads to
\begin{eqnarray}\label{app:46}
\hat{W}^{(inter)}_{nres}|X_{\textbf{K},\nu_0;\mu,S=0}\rangle=
\hat{U}^\dag_{nres}|X_{\textbf{K},\nu_0;\mu,S=0}\rangle 
\\
= -\sum_{\textbf{Q}_1,\nu_1,\mu_1}
\frac{\Lambda^*_{ \nu_1}}{{\sqrt{\hbar\omega_{\textbf{Q}_1}}}}
 \sum_{\lambda_1=(X,Y)} (\textbf{e}_{\lambda_1,\textbf{Q}_1}\cdot\textbf{e}_{\mu_1})
 |f_1\rangle
 \nonumber
\end{eqnarray}
where the intermediate state $ |f_1\rangle$ is made of a ($\textbf{Q}_1,\lambda_1)$ photon plus two spin-singlet excitons (see Fig.~S\ref{fig:S5}(b,c))
\begin{equation}
|f_1\rangle=
\hat{\alpha}^\dag_{_{\textbf{Q}_1,\lambda_1}}
|v\rangle
\otimes
\hat{B}^\dag_{-\textbf{Q}_1,\nu_1;\mu_1,S=0}
\hat{B}^\dag_{\textbf{K},\nu_0;\mu,S=0}
|v\rangle\label{app:47}
\end{equation}
with a similar intermediate state $\langle f_2|$   for $\langle X_{\textbf{K},\nu_0;\mu',S=0}|\hat{U}_{nres}$.

The $\langle f_2|f_1\rangle$ scalar product imposes the two photons to be identical, $(\textbf{Q}_1=\textbf{Q}_2,\lambda_1=\lambda_2)$. The remaining exciton part gives two terms
\begin{eqnarray}
\langle v| \hat{B}_0 \hat{B}_2 \hat{B}_1^\dag \hat{B}_0^\dag | v \rangle\simeq \delta_{0,0}\delta_{2,1}+\delta_{0,1}\delta_{2,0}
\end{eqnarray}
plus a fermion exchange term which scales as the inverse of the sample volume, negligible in front of the delta terms \cite{appCombescot_book}. The direct term in $\delta_{0,0}\delta_{2,1}$, shown in Fig.~S\ref{fig:S5}(b), does not contribute to the change in the exciton energy because in this diagram the virtual photon is not connected to the exciton. We are left with the indirect term in $\delta_{0,1}\delta_{2,0}$, shown in Fig.~S\ref{fig:S5}(c), which imposes $-\textbf{Q}_1=\textbf{K}=-\textbf{Q}_2$.

So, we end with the contribution from nonresonant photon-electron interaction given by 
\begin{eqnarray}
\Delta_{nres}^{(\mu',\mu)} = \frac{|\Lambda_{\nu_0}|^2}{\hbar\omega_{-\textbf{K}}}\frac{1}{E_{\textbf{K},\nu_0}-(\hbar\omega_{-\textbf{K}}+2E_{\textbf{K},\nu_0})}
\nonumber
\\
 \sum_{\lambda=(X,Y)} \left(  \textbf{e}_{\lambda,\textbf{K}}\cdot\textbf{e}_  {\mu'} \right)
\left(  \textbf{e}_{\lambda,\textbf{K}}\cdot\textbf{e}_\mu \right)
\end{eqnarray}

$\bullet$\textit{ Interband photon-electron contribution}

By adding contributions from the resonant and nonresonant interband photon-electron couplings, $\Delta_{ph}^{(\mu',\mu)}=\Delta_{res}^{(\mu',\mu)}+\Delta_{nres}^{(\mu',\mu)}$, we end, since $\omega_\textbf{K}=\omega_{-\textbf{K}}$, with
\begin{eqnarray}
\Delta_{ph}^{(\mu',\mu)} =  \frac{|\Lambda_{\nu_0}|^2}{\hbar\omega_\textbf{K}} \left( \frac{1}{E_{\textbf{K},\nu_0}-\hbar\omega_\textbf{K}} + \frac{1}{-E_{\textbf{K},\nu_0}-\hbar\omega_\textbf{K}} \right)\nonumber\\
   \sum_{\lambda=(X,Y)} \left(  \textbf{e}_{\lambda,\textbf{K}}\cdot\textbf{e}_  {\mu'} \right)
\left(  \textbf{e}_{\lambda,\textbf{K}}\cdot\textbf{e}_\mu \right)\hspace{1cm}\label{app:51}\\
= 2 \, \frac{|\Lambda_{\nu_0}|^2}{E_{\textbf{K},\nu_0}^2-\hbar^2\omega_\textbf{K}^2} \sum_{\lambda=(X,Y)} \left(  \textbf{e}_{\lambda,\textbf{K}}\cdot\textbf{e}_  {\mu'} \right)
\left(  \textbf{e}_{\lambda,\textbf{K}}\cdot\textbf{e}_\mu \right)\nonumber
\end{eqnarray}
with    $E_{\textbf{K},\nu_0}^2-\hbar^2\omega_\textbf{K}^2\simeq E_{\textbf{K},\nu_0}^2\simeq E_{gap}^2$.

\subsection {Total shift due to the electromagnetic field}

The last step is to add the interband Coulomb contribution given in Eq.~(\ref{app:42}) to the interband electron-photon contribution given in Eq.~(\ref{app:51}). This is done by noting that $\textbf{K}/K=\textbf{e}_{Z,\textbf{K}}$ and $(\textbf{e}_{X,\textbf{K}},\textbf{e}_{Y,\textbf{K}})$ form an orthonormal set; so,
\begin{eqnarray}
&{}&\delta_{\mu',\mu}=\textbf{e}_{\mu'} \cdot \textbf{e}_{\mu}
\\
&=&\sum_{\lambda'=(X,Y,Z)}  ( \textbf{e}_{\mu'} \cdot \textbf{e}_{\lambda',\textbf{K}})  \textbf{e}_{\lambda',\textbf{K}}
\cdot
\sum_{\lambda=(X,Y,Z)}  ( \textbf{e}_{\mu} \cdot \textbf{e}_{\lambda,\textbf{K}})  \textbf{e}_{\lambda,\textbf{K}}
\nonumber\\
&=&\sum_{\lambda=(X,Y)} \left( \textbf{e}_  {\mu'} \cdot \textbf{e}_{\lambda,\textbf{K}}\right)
\left(  \textbf{e}_{\lambda,\textbf{K}}\cdot\textbf{e}_\mu \right)
\nonumber
\end{eqnarray}
Using this result, we readily get the energy shift of the ground-state exciton in a spin-singlet state as
\begin{equation}
\Delta_{Coul}^{(\mu',\mu)}+\Delta_{ph}^{(\mu',\mu)} \simeq  \delta_{\mu',\mu}\,\frac{2|\Lambda_{\nu_0}|^2}{E_{gap}^2} 
\end{equation}
 Since dark excitons, made of same-spin carriers, are not coupled to the electromagnetic field, the above shift just corresponds to the energy splitting between bright and dark excitons. When expressed in terms of the exciton-photon Rabi couplings given in Eq.~(\ref{app:41}), we find that the bright-dark splitting $\Delta_{BD}$ scales as
\begin{equation}
\Delta_{BD} \propto \, \frac{|\Omega_{ph-X}|^2}{E_{gap}}
\end{equation}


\begin{thebibliography}{99}
 
 \bibitem{Knox} R. S. Knox, \textit{Theory of Excitons} (Academic Press, 1963).
 \bibitem{CHO} K. Cho, \textit{``Excitons'' Topics in current physics} Vol.14, Springer-Verlag (1979). 
 
 \bibitem{Combescot_book} M. Combescot and S.-Y. Shiau, \textit{Excitons and Cooper Pairs: two composite bosons in many-body physics} (Oxford Univ. Press, 2015).
 
 \bibitem{Phys_Rep} M. Combescot, O. Betbeder-Matibet, and F. Dubin, Phys. Rep. \textbf{463}, 215 (2008).
 
 \bibitem{Combescot_2015} M. Combescot, R. Combescot, M. Alloing, and F. Dubin, Phys. Rev. Lett. \textbf{114}, 090401 (2015).
 
 \bibitem{Bimberg_1979} W. Ekardt, K. L\"{o}sch, and D. Bimberg, Phys. Rev B \textbf{20}, 3303 (1979).

\bibitem{Blackwood_94} E. Blackwood, M. J. Snelling, R. T. Harley, S. R. Andrews, and C. T. B. Foxon, 
Phys. Rev. B \textbf{50}, 14246 (1994).

\bibitem{Amand_97} T. Amand \textit{et al.}, Phys. Rev. Lett. \textbf{78}, 1355 (1997).

\bibitem{Lavallard}  I. V. Mashkov \textit{et al.}, Phys. Rev. B \textbf{55}, 13761 (1997).

\bibitem{Combescot_2007} M. Combescot, O. Betbeder-Matibet, and R. Combescot, Phys. Rev. Lett. \textbf{99}, 176403 (2007).

\bibitem{Leunberger} M. Combescot, and M. Leunberger, Sol. Stat. Commun. \textbf{149}, 13 (2009).

\bibitem{Combescot_ROPP} M. Combescot, R. Combescot, and F. Dubin, Rep. Prog. Phys. \textbf{80}, 066401 (2017).

\bibitem{Sean2019} S.-Y. Shiau, and M. Combescot, Phys. Rev. Lett. \textbf{123}, 097401 (2019).

\bibitem{Kis_2019}D. Unuchek, A. Ciarrocchi, A. Avsar, Z. Sun, K. Watanabe, T. Taniguchi, and A. Kis, Nat. Nanotechnol. \textbf{14},  1104 (2019). 


\bibitem{Gershoni_2016} I. Schwarz \textit{et al.}, Science  \textbf{354}, 434 (2016).

\bibitem{Xu_2018} P. Rivera, H. Yu, K. L. Seyler, N. P. Wilson, W. Yao, and X. Xu, Nat. Nanotechnol. \textbf{13}, 1004 (2018).

\bibitem{Robert_2020} C. Robert \textit{et al.}, Nat. Commun. \textbf{11}, 4037 (2020). 

\bibitem{Pikus_1971} G. E. Pikus, G. L. Bir, Soc. Phys. JETP \textbf{33}, 108 (1971).

\bibitem{Sham_93} M. Z. Maialle, E. A. de Andrada e Silva, and L. J. Sham, Phys. Rev. B \textbf{47}, 15776 (1993).

\bibitem{Fu_99} H. Fu, L.-W. Wang, and A. Zunger, Phys. Rev. B \textbf{59}, 5568 (1999).
\bibitem{Luo2009} J.-W. Luo, G. Bester, and A. Zunger, New J. Phys. \textbf{11}, 123024 (2009). 
 
 \bibitem{Lamb_47} W. E. Lamb, and R. C. Retherford, Phys. Rev. \textbf{72}, 241 (1947).
  
\bibitem{Bethe_47} H. A. Bethe, Phys. Rev. \textbf{72}, 339 (1947).

\bibitem{Cohen_Atom_Photon} C. Cohen-Tannoudji, J. Dupont Roc, and G. Grynberg, \textit{Atom-Photon Interactions: basic processes and applications} (Wiley-VCH, 1998). 

\bibitem{note1} The conduction band splitting of transition metal dichalcogenides, induced by spin-orbit interaction, also produces a bright-dark splitting which is usually small compared to the splitting induced by electromagnetic quantum fluctuations.

\bibitem{SM} See online supplements.

\bibitem{SOcase} By working with spin and linear polarizations, instead of  spin-orbit eigenstates and circular polarizations, the interplay between $ \textbf{e}_\mu$ and \textbf{K} in terms of ``longitudinal'' and ``transverse'' contributions, is easy to pin down. The splitting prefactors for spin-orbit eigenstates can be obtained by writing these states in terms of ($\mu,s$) hole states.

\bibitem{dielectric} We have derived the Coulomb and electron-photon vertices in vacuum. Yet, semiconductors have a dielectric constant $\epsilon_{sc}$ that comes from dressing the Coulomb interaction through interband Coulomb processes. This goes with replacing  $e^2$ by $e^2/\epsilon_{sc}$ in the parameter $\Lambda$ of Eq.~(\ref{4}).

\bibitem{Bastard} G. Bastard, \textit{Wave Mechanics Applied to Semiconductor Heterostructures} (Les Editions de Physique, 1990).

\bibitem{High_2012} A.A. High \textit{et al.}, Nature \textbf{483}, 584 (2012).

\bibitem{Alloing_2014} M. Alloing \textit{et al.}, Europhys. Lett. \textbf{107}, 10012 (2014).

\bibitem{Rapaport_trap} Y. Shilo \textit{et al.}, Nat. Commun. \textbf{4}, 2335 (2013).

\bibitem{Beian_2017} M. Beian \textit{et al.}, Europhys. Lett. \textbf{119}, 37004 (2017). 

\bibitem{Anankine_2017} R. Anankine \textit{et al.}, Phys. Rev. Lett. \textbf{118}, 127402 (2017).

\bibitem{Dang_2019} S. Dang \textit{et al.}, Phys. Rev. Lett. \textbf{122}, 117402 (2019).

\bibitem{MFF} In good bulk samples, bright excitons face the fact that they form polaritons.


\bibitem{Deveaud_91}B. Deveaud, F. Clerot, N. Roy, K. Satzke, B. Sermage, and D. S. Katzer,
Phys. Rev. Lett. \textbf{67}, 2355 (1991).

\bibitem{Sivalertpron_2012} K. Sivalertporn, L. Mouchliadis, A. L. Ivanov, R. Philp, and E. A. Muljarov,
Phys. Rev. B \textbf{85}, 045207 (2012).



\end{thebibliography}

\begin{thebibliography}{99}
 


\bibitem{appCombescot_book} M. Combescot and S.Y. Shiau, \textit{Excitons and Cooper Pairs: Two Composite Bosons in Many-Body Physics} (Oxford. Univ. Press, 2015).




\end{thebibliography}
\end{document}